# Extraordinary transverse spin: Hidden vorticity of the energy flow and momentum distributions in propagating light fields

A. Y. Bekshaev

Physics Research Institute, I.I. Mechnikov National University,
Dvorianska 2, 65082, Odessa, Ukraine,
e-mail: bekshaev@onu.edu.ua

**Abstract**

Spatially inhomogeneous fields of electromagnetic guided modes exhibit a complex of extraordinary dynamical properties such as the polarization-dependent transverse momentum, helicity-independent transverse spin, spin-associated non-reciprocity and unidirectional propagation, etc. Recently, the remarkable relationship has been established between the spin and propagation features of such fields, expressed through the spin–momentum equations [Proc. Natl. Acad. Sci. 118 (2021) e2018816118] connecting the wave spin with the curl of momentum. Here, the meaning, limitations and specific forms of this correspondence are further investigated, involving the physically transparent and consistent examples of paraxial light fields, plane-wave superpositions and evanescent waves. The conclusion is inferred that the spin–momentum equation is an attribute of guided waves with well defined direction of propagation, and it unites the helicity-independent "extraordinary" transverse spin with the spatially-inhomogeneous longitudinal field momentum (energy flow) density. Physical analogies with the layered hydrodynamic flows and possible generalizations for other wave fields are discussed. The results can be useful in optical trapping, manipulation and the data processing techniques.

## 1. Introduction

During the recent years, light fields with strong and controllable spatial inhomogeneity become a hot topic of modern optics [1–6]. Such fields find impressive applications in optical nano-probing, micromanipulation, optical control of materials, data encoding and processing. Moreover, they play an exclusive role in the basic research projects aimed at the most fundamental problems of the light-matter interaction, information capacity and nature of a photon, mechanisms of the light emission and absorption, studies of the quantum-classic boundary, weak measurements, etc. [1,5,7].

In the course of this vibrant and fruitful activity, especial attention is paid to the dynamical properties of such fields associated with the spatial distributions of their energy, momentum and angular momentum [6,7,8–10]. In particular, a complex structure of electromagnetic momentum has been revealed: it consists of the orbital (canonical) and spin (virtual) contributions, reflecting the roles of spatial (orbital) and polarization (spin) degrees of freedom [6,5,10]; likewise, the electromagnetic spin can also be decomposed into physically meaningful parts, differently related to the field polarization. It is found that the polarization-dependent "spin momentum" (SM) $\mathbf{p}_S$ shows an "extraordinary" behavior: it can be directed orthogonally to the predominant direction of light propagation [11–16]. Quite similarly, the "transverse" orientation is typical for the helicity-independent "extraordinary spin" (ES) $\mathbf{s}_E$ associated with the rotation of the field vectors in the longitudinal plane [12–15].

Such impressive features strongly enrich the common ideas on the electromagnetic spin and momentum and disclose the deep and far-reaching interrelations between the dynamical characteristics of wave fields in general. The natural spread of this approach to fields of other nature has contributed, e.g., to discovery of the spin of acoustic or elastic waves [17–20]. A particularly remarkable fact is that the SM of light is determined by a curl of the electromagnetic spin distribution [5,10]; in turn, the ES of spatially localized and guided light waves shows the strict coupling to the momentum inhomogeneity and the direction of light propagation ("spin–momentum locking") [15,21–23].

Further development of these ideas has led to the search of more distinct and regular relations between the ES and momentum of light. In this way, P. Shi et al. [24] have noticed that the ES can be associated with the curl of the "usual" (not extraordinary) momentum distribution and thus a remarkable "duality" between the SM and ES takes place:

$$\mathbf{p}_S \propto \nabla \times \mathbf{s}\,, \tag{1}$$

$$\mathbf{s}_E \propto \nabla \times \mathbf{p} \tag{2}$$

where $\mathbf{p}$ and $\mathbf{s}$ are the electromagnetic momentum and spin densities, and subscripts "$S$" and "$E$" denote the "spin" and "extraordinary" constituents. (It is tempting to call $\mathbf{p}_S$ "extraordinary momentum" in view of its above-mentioned extraordinary properties but we prefer to keep the traditional wording supported by the fact that $\mathbf{p}_S$ is a part of the quite ordinary Poynting momentum, see Eqs. (4) and (5) below). Note that Eq. (1) is correct by the definition while (2) follows from additional presumptions, not always clear. Based on Eqs. (1) and (2), authors of [24] propose a set of spin–momentum equations valid for electromagnetic guided waves, which are analogous to the Maxwell's equations. However, despite the wittiness and beauty of this approach, it is not absolutely valid, and the limits of its applicability as well as the exact meaning of the separate terms of the "spin–momentum equations", are not clearly specified so far.

In this paper, we inspect additional examples of structured light fields and study the applicability of Eq. (2) for them, paying especial attention to the meaning of physical quantities it unites. We show that, with slight modification, the spin–momentum relations are rather universal and can be applied to paraxial light beams and even to simple superpositions of plane waves, provided that the meanings of $\mathbf{p}$ and $\mathbf{s}_E$ are properly understood. Additionally, we find specific refinements of the spin–momentum relations with account for the medium properties and the field character. Finally, we discuss the general ES properties and their interpretation based on mechanical analogies with the vorticity of the layered flow in a continuous medium [25].

## 1. Basic equations

We consider monochromatic electromagnetic field in a lossless medium with permittivity $\varepsilon$ and permeability $\mu$ (possibly, frequency-dependent), and use the Gaussian system of units. The space coordinates are given by 3D radius-vector $\mathbf{R} = \mathbf{r} + \mathbf{e}_z z$, where the longitudinal coordinate $z$ is associated with the predominant direction of the beam propagation (whose existence is supposed) and $\mathbf{r} = \mathbf{e}_x x + \mathbf{e}_y y$ is the transverse radius-vector; $\mathbf{e}_x$, $\mathbf{e}_y$ and $\mathbf{e}_z$ are the unit vectors of the Cartesian frame. The real electric and magnetic fields, $\boldsymbol{E}(\mathbf{R},t)$ and $\boldsymbol{H}(\mathbf{R},t)$, oscillating with frequency $\omega$, can be described by the complex-valued phasors:

$$\boldsymbol{E}(\mathbf{R},t) = \mathrm{Re}\big[\mathbf{E}(\mathbf{R})\exp(-i\omega t)\big], \quad \boldsymbol{H}(\mathbf{R},t) = \mathrm{Re}\big[\mathbf{H}(\mathbf{R})\exp(-i\omega t)\big].$$

The field energy averaged over the period of oscillations is distributed with the volume density [26,27]

$$w = \frac{g}{2}\left(\tilde{\varepsilon}|\mathbf{E}|^2 + \tilde{\mu}|\mathbf{H}|^2\right)$$

where

$$\tilde{\varepsilon} = \varepsilon + \omega\frac{d\varepsilon}{d\omega}\,, \quad \tilde{\mu} = \mu + \omega\frac{d\mu}{d\omega} \tag{3}$$

are the dispersion-modified permittivity and permeability of the medium. The energy flow density is given by the time-averaged Poynting vector $\mathbf{S}$ associated with the kinetic Abraham momentum $\mathbf{p}^A$ [27,8]:

$$\mathbf{S} = c^2 \mathbf{p}^A, \quad \mathbf{p}^A = \frac{g}{c} \mathrm{Re}\left[\mathbf{E}^* \times \mathbf{H}\right]. \tag{4}$$

Here $g = (8\pi)^{-1}$ and $c$ is the velocity of light in vacuum. The "true" kinetic electromagnetic momentum density in the medium is described by the dispersion-modified Minkowski expressions [8,9]

$$\mathbf{p} = \mathbf{p}_S + \mathbf{p}_O \tag{5}$$

where

$$\mathbf{p}_S = \mathbf{p}_S^e + \mathbf{p}_S^m = \frac{g}{4\omega} \mathrm{Im}\left[\nabla \times \left(\tilde{\varepsilon}\mathbf{E}^* \times \mathbf{E} + \tilde{\mu}\mathbf{H}^* \times \mathbf{H}\right)\right], \tag{6}$$

is the spin ("extraordinary") momentum,

$$\mathbf{p}_O = \mathbf{p}_O^e + \mathbf{p}_O^m = \frac{g}{2\omega} \mathrm{Im}\left[\tilde{\varepsilon}\mathbf{E}^* \cdot (\nabla)\mathbf{E} + \tilde{\mu}\mathbf{H}^* \cdot (\nabla)\mathbf{H}\right] \tag{7}$$

represents the orbital (canonical) constituent, and the superscripts "$e$" and "$m$" in Eqs. (6) – (8) denote the electric and magnetic contributions. The electromagnetic spin density is determined as

$$\mathbf{s} = \mathbf{s}^e + \mathbf{s}^m = \frac{g}{2\omega} \mathrm{Im}\left(\tilde{\varepsilon}\mathbf{E}^* \times \mathbf{E} + \tilde{\mu}\mathbf{H}^* \times \mathbf{H}\right). \tag{8}$$

Importantly,

$$\mathbf{p}_S = \frac{1}{2}\nabla \times \mathbf{s}, \tag{9}$$

that is, the SM originates from inhomogeneity of the optical spin (cf. Eq. (1)).

## 2. Paraxial beams

For simplicity, we start with consideration of paraxial beams in spatially homogeneous media where the $z$-axis is the physically selected longitudinal direction. In this case, the electric and magnetic vectors of the optical field can be expressed as [6,10]

$$\mathbf{E} = \mathbf{E}_\perp + \mathbf{e}_z E_z = \left[\mathbf{u} + \frac{i}{k}\mathbf{e}_z (\nabla_\perp \cdot \mathbf{u})\right] e^{ikz}, \tag{10}$$

$$\mathbf{H} = \mathbf{H}_\perp + \mathbf{e}_z H_z = \left[(\mathbf{e}_z \times \mathbf{u}) + \frac{i}{k}\mathbf{e}_z (\nabla_\perp \cdot (\mathbf{e}_z \times \mathbf{u}))\right] e^{ikz}. \tag{11}$$

where $k = n(\omega/c)$ is the wavenumber, $n = \sqrt{\varepsilon\mu}$ is the refractive index, and $\mathbf{u}(\mathbf{r}, z)$ is the slowly varying complex amplitude – the transverse vector function given by

$$\mathbf{u} = \mathbf{e}_x u_x + \mathbf{e}_y u_y = \mathbf{e}_+ u_+ + \mathbf{e}_- u_- \tag{12}$$

with

$$\mathbf{e}_+ = \frac{1}{\sqrt{2}}\left(\mathbf{e}_x + i\mathbf{e}_y\right), \quad \mathbf{e}_- = \mathbf{e}_+^* = \frac{1}{\sqrt{2}}\left(\mathbf{e}_x - i\mathbf{e}_y\right), \tag{13}$$

$$u_x = \frac{u_+ + u_-}{\sqrt{2}}, \quad u_y = i\frac{u_+ - u_-}{\sqrt{2}}. \tag{14}$$

The complex basis (13) is naturally associated with the complex transverse coordinates

$$\xi = \frac{x + iy}{\sqrt{2}}, \quad \eta = \xi^* = \frac{x - iy}{\sqrt{2}}, \tag{15}$$

through which the transverse gradient is represented as

$$\nabla_\perp = \mathbf{e}_x \frac{\partial}{\partial x} + \mathbf{e}_y \frac{\partial}{\partial y} = \mathbf{e}_+ \frac{\partial}{\partial \xi} + \mathbf{e}_- \frac{\partial}{\partial \eta}. \tag{16}$$

Equations (12) – (14) supply the complex amplitude representations in bases of circular ($u_\pm$) or linear ($u_{x,y}$) polarizations, which are equally admissible in the paraxial approximation. The complex amplitude components obey the parabolic equation [6,10]

$$i \frac{\partial u_\sigma}{\partial z} = -\frac{1}{2k} \nabla_\perp^2 u_\sigma \tag{17}$$

where $\sigma = \pm 1$ or $\sigma = x, y$ according to the basis accepted. The main (first) terms in the right-hand sides of (10) and (11) describe the transverse field components, whereas the longitudinal components (second terms) are of the relative order $\delta = (kb)^{-1}$ in magnitude, with $b$ being the characteristic transverse scale of the complex amplitude distribution $\mathbf{u}(\mathbf{r}, z)$. The quantity $\delta$ is the small parameter of the paraxial approximation [5,10].

By using Eqs. (5) – (15) we can find the momentum density constituents in the first paraxial order. For convenience, we start with the electric contributions, which can be recast in the form

$$\mathbf{p}^e = \mathbf{p}_\parallel^e + \mathbf{p}_{O\perp}^e + \mathbf{p}_{S\perp}^e \tag{18}$$

where

$$\mathbf{p}_\parallel^e = \frac{gk}{2\omega} \tilde{\varepsilon} \left( |u_+|^2 + |u_-|^2 \right) \mathbf{e}_z \tag{19}$$

is the longitudinal (main) component, while the other terms describe the transverse momentum constituents being of the first order in $\delta$ with respect to (19):

$$\mathbf{p}_{O\perp}^e = \frac{g}{2\omega} \tilde{\varepsilon} \left( |u_+|^2 \nabla_\perp \varphi_+ + |u_-|^2 \nabla_\perp \varphi_- \right) \tag{20}$$

is the transverse orbital momentum, and

$$\mathbf{p}_{S\perp}^e = \frac{g}{4\omega} \tilde{\varepsilon} \, \mathrm{Im} \left[ \nabla_\perp \times \left( \mathbf{u}^* \times \mathbf{u} \right) \right] = \frac{g}{4\omega} \tilde{\varepsilon} \nabla_\perp \times \left[ \mathbf{e}_z \left( |u_+|^2 - |u_-|^2 \right) \right] \tag{21}$$

is the SM. Corresponding magnetic contributions can be derived similarly and follow from Eqs. (18) – (21) according to the common rule

$$\mathbf{p}_\alpha^m = \frac{\tilde{\mu} \varepsilon}{\mu \tilde{\varepsilon}} \mathbf{p}_\alpha^e \tag{22}$$

where $\alpha$ is an arbitrary subscript of Eq. (18) (or no subscript). Consequently, components of the total electromagnetic momentum density in a paraxial field can be presented in the form

$$\mathbf{p}_\alpha = \left( 1 + \frac{\tilde{\mu} \varepsilon}{\mu \tilde{\varepsilon}} \right) \mathbf{p}_\alpha^e, \quad \mathbf{p} = \left( 1 + \frac{\tilde{\mu} \varepsilon}{\mu \tilde{\varepsilon}} \right) \mathbf{p}^e.$$

Note that Eq. (21) directly relates the SM with the field helicity [6,7] that in paraxial fields (within the first order in $\delta$) can be expressed as [6,8]

$$K = -\frac{g}{2\omega} \left( \frac{\tilde{\varepsilon}}{\varepsilon} + \frac{\tilde{\mu}}{\mu} \right) \mathrm{Im} \left( \mathbf{E}^* \cdot \mathbf{H} \right) = \frac{g}{\omega} \left( \frac{\tilde{\varepsilon}}{\varepsilon} + \frac{\tilde{\mu}}{\mu} \right) \left( |u_+|^2 - |u_-|^2 \right). \tag{23}$$

The spin density admits the similar representation. Its electric contribution can be presented as

$$\mathbf{s}^e = \mathbf{s}_{H\parallel}^e + \mathbf{s}_{H\perp}^e + \mathbf{s}_{E\perp}^e + \mathbf{s}_{I\perp}^e \tag{24}$$

where

$$\mathbf{s}_{H\parallel}^e = \frac{g}{2\omega} \tilde{\varepsilon} \left( |u_+|^2 - |u_-|^2 \right) \mathbf{e}_z, \tag{25}$$

$$\mathbf{s}_{H\perp}^e = \frac{g}{2\omega} \tilde{\varepsilon} \left( |u_+|^2 \nabla_\perp \varphi_+ - |u_-|^2 \nabla_\perp \varphi_- \right) \tag{26}$$

are the longitudinal and transverse components of the usual spin directly associated with the field helicity and vanishing in the linearly polarized or scalar fields. On the contrary, the third term of (24) is helicity-independent and represents the transverse ES [6]:

$$\mathbf{s}_{E\perp}^{e} = \frac{g}{4\omega k}\left(\mathbf{e}_x \frac{\partial}{\partial y} - \mathbf{e}_y \frac{\partial}{\partial x}\right)\left(\left|u_+\right|^2 + \left|u_-\right|^2\right) = -\frac{g}{4\omega k}\mathrm{Im}\left(\mathbf{e}_+ \frac{\partial}{\partial \xi} - \mathbf{e}_- \frac{\partial}{\partial \eta}\right)\left(\left|u_+\right|^2 + \left|u_-\right|^2\right). \tag{27}$$

The fourth term of (24) describes the specific contribution due to interference between the opposite helicity components,

$$\mathbf{s}_{I\perp}^{e} = -\frac{g}{2\omega k}\mathrm{Im}\left(\mathbf{e}_+ \frac{\partial}{\partial \eta}\left(u_+ u_-^*\right) - \mathbf{e}_- \frac{\partial}{\partial \xi}\left(u_- u_+^*\right)\right). \tag{28}$$

As to the magnetic contributions, the rule similar to (22),

$$\mathbf{s}_{\alpha}^{m} = \frac{\tilde{\mu}\varepsilon}{\mu\tilde{\varepsilon}}\mathbf{s}_{\alpha}^{e}, \tag{29}$$

still holds for the three summands of Eq. (24) but for the last one the sign is reversed: $\mathbf{s}_{I\perp}^{m} = -\frac{\tilde{\mu}\varepsilon}{\mu\tilde{\varepsilon}}\mathbf{s}_{I\perp}^{e}$. Accordingly, the total spin density of a paraxial field can be reduced to

$$\mathbf{s} = \left(1 + \frac{\tilde{\mu}\varepsilon}{\mu\tilde{\varepsilon}}\right)\left(\mathbf{s}_{H\parallel}^{e} + \mathbf{s}_{H\perp}^{e} + \mathbf{s}_{E\perp}^{e}\right) + \left(1 - \frac{\tilde{\mu}\varepsilon}{\mu\tilde{\varepsilon}}\right)\mathbf{s}_{I\perp}^{e}. \tag{30}$$

Note that in non-dispersive media, the first three contributions of (24) are simply doubled in (30) while the interference contribution associated with $\mathbf{s}_{I\perp}^{e}$ vanishes.

Our main interest is focused on the ES (27) which is independent of the field helicity (23); moreover, it does not depend on the field polarization in the transverse plane (is completely determined by the "total" intensity $\left|u_+\right|^2 + \left|u_-\right|^2$, regardless of how is it divided between the polarization components). This spin appears due to the field vector rotation in the longitudinal plane and its analogues were recently discovered in a series of structured optical fields [6,10–16] as well as in structured fields of other physical natures [17–19]. By comparison with Eqs. (18), (19) one can easily persuade that the ES obeys the relation

$$\mathbf{s}_{E}^{e,m} = \frac{1}{2k^2}\nabla \times \mathbf{p}_{\parallel}^{e,m}, \quad \mathbf{s}_E = \mathbf{s}_E^e + \mathbf{s}_E^m = \frac{1}{2k^2}\nabla \times \mathbf{p}_{\parallel}. \tag{31}$$

This equation confirms the idea of Eq. (2) for general paraxial beams and shows that the spin $\mathbf{s}_E$ originates from the spatial inhomogeneity of the field momentum and/or energy – just like the SM (9) originates from the inhomogeneity of the spin. It completely coincides with the "spin-momentum equation" of [24] and generalizes the latter to the wide variety of optical fields that can be described by the paraxial approximation [6,10].

## 3. Two-wave interference field

Another instructive example of structured light is the field formed by interference of two plane waves with equal amplitudes and arbitrary polarizations [13] (see Fig. 1). We consider plane waves with the wavevectors $\mathbf{k}_{1,2} = k\left(\mathbf{e}_z \cos\gamma \pm \mathbf{e}_x \sin\gamma\right)$, possessing the electric and magnetic vectors with complex amplitudes

$$\mathbf{E}_{1,2} = \frac{A}{\sqrt{1 + \left|m_1\right|^2}}\left(\mathbf{e}_x \cos\gamma + \mathbf{e}_y m_{1,2} \mp \mathbf{e}_z \sin\gamma\right)e^{i\Phi_{1,2}}, \tag{32}$$

$$\mathbf{H}_{1,2} = \sqrt{\frac{\varepsilon}{\mu}}\frac{A}{\sqrt{1 + \left|m_1\right|^2}}\left(-\mathbf{e}_x m_{1,2}\cos\gamma + \mathbf{e}_y \pm \mathbf{e}_z m_{1,2}\sin\gamma\right)e^{i\Phi_{1,2}}. \tag{33}$$

Here, both waves have equal amplitude factors $A$, their polarizations are characterized by complex parameters $m_{1,2}$, and $\Phi_{1,2} = k\left(z\cos\gamma \pm x\sin\gamma\right)$ are the wave phases.

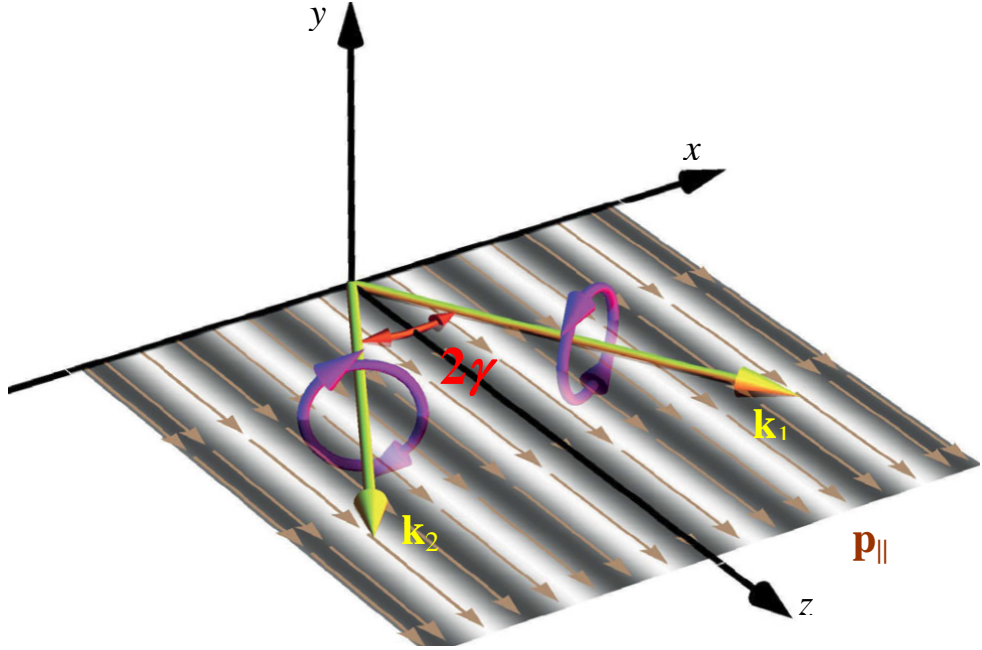

Fig. 1. Interference of two polarized plane waves of Eqs. (32), (33); their directions are specified by the wavevectors $\mathbf{k}_1$ and $\mathbf{k}_2$ lying in the plane $(x, z)$ with an angle $2\gamma$ between them. The wave polarizations are arbitrary but here the case of opposite circular polarizations is shown. The gray-scale plot represents the distribution of the electric part of the orbital momentum density (34), the in-plane brown arrows show its direction.

In the interference field $\mathbf{E}_1 + \mathbf{E}_2$, $\mathbf{H}_1 + \mathbf{H}_2$, the electromagnetic momentum and spin densities can be easily determined from (5) – (8) and (32), (33). Moreover, the momentum and spin components (in the Abraham definition) were explicitly calculated in the Supplemental Material of [13]. So, after obvious transformations with account for the medium dispersion [8,9] and omitting the common factor $\varepsilon g |A|^2$, one obtains

$$\mathbf{p}_O^e = n\frac{\tilde{\varepsilon}}{\varepsilon}\frac{\cos\gamma}{cN}\mathbf{e}_z\left\{N + \mathrm{Re}\left[\left(\cos 2\gamma + m_1 m_2^*\right)e^{i\Phi}\right]\right\}, \tag{34}$$

$$\mathbf{p}_S^e = n\frac{\tilde{\varepsilon}}{\varepsilon}\frac{\sin 2\gamma}{cN}\left\{\mathbf{e}_z\sin\gamma\,\mathrm{Re}\left(e^{i\Phi}\right) - \mathbf{e}_y\frac{1}{2}\mathrm{Re}\left[\left(m_1 - m_2^*\right)e^{i\Phi}\right]\right\}; \tag{35}$$

$$\mathbf{p}_O^m = n\frac{\tilde{\mu}}{\mu}\frac{\cos\gamma}{cN}\mathbf{e}_z\left\{N + \mathrm{Re}\left[\left(m_1 m_2^*\cos 2\gamma + 1\right)e^{i\Phi}\right]\right\}, \tag{36}$$

$$\mathbf{p}_S^m = n\frac{\tilde{\mu}}{\mu}\frac{\sin 2\gamma}{cN}\left\{\mathbf{e}_z\sin\gamma\,\mathrm{Re}\left(m_1 m_2^* e^{i\Phi}\right) - \mathbf{e}_y\frac{1}{2}\mathrm{Re}\left[\left(m_1 - m_2^*\right)e^{i\Phi}\right]\right\} \tag{37}$$

where $N = \sqrt{1 + |m_1|^2}\sqrt{1 + |m_2|^2}$, $\Phi = 2kx\sin\gamma$.

In this case, the orbital momentum (7), expressed by Eqs. (34) and (36), is completely longitudinal ($\sim\mathbf{e}_z$) while the SM (6) constituents (Eqs. (35), (37)) contain both longitudinal ($\sim\mathbf{e}_z$) and transverse ($\sim \mathbf{e}_y$) components. The spin density also possesses all three spatial components [13] but here we consider only the helicity-independent $y$-directed ES contributions:

$$\mathbf{s}_E^e = \frac{\tilde{\varepsilon}}{\varepsilon}\frac{\sin 2\gamma}{\omega N}\mathbf{e}_y \,\text{Im}\left(e^{i\Phi}\right), \quad \mathbf{s}_E^m = \frac{\tilde{\mu}}{\mu}\frac{\sin 2\gamma}{\omega N}\mathbf{e}_y \,\text{Im}\left(m_1 m_2^* e^{i\Phi}\right). \tag{38}$$

In agreement to Eqs. (34) and (35),

$$\mathbf{p}_\parallel^e = \mathbf{p}_{O\parallel}^e + \mathbf{p}_{S\parallel}^e = n\frac{\tilde{\varepsilon}}{\varepsilon}\frac{\cos\gamma}{cN}\mathbf{e}_z\left\{N + \text{Re}\left[\left(1 + m_1 m_2^*\right)e^{i\Phi}\right]\right\}, \tag{39}$$

$$\mathbf{p}_\parallel^m = \mathbf{p}_{O\parallel}^m + \mathbf{p}_{S\parallel}^m = \frac{\tilde{\mu}}{\mu}\frac{\varepsilon}{\tilde{\varepsilon}}\mathbf{p}_\parallel^e \tag{40}$$

(cf. Eqs. (22) and (29)). Confrontation of Eqs. (38) and (39), (40) leads to a conclusion that here, again, relations similar to (2) and (31) take place but they look differently for the electric and magnetic contributions:

$$\mathbf{s}_E^e = \frac{1}{k^2}\frac{\text{Im}\left(e^{i\Phi}\right)}{\text{Im}\left[\left(1 + m_1 m_2^*\right)e^{i\Phi}\right]}\nabla\times\mathbf{p}_\parallel^e, \quad \mathbf{s}_E^m = \frac{1}{k^2}\frac{\text{Im}\left(m_1 m_2^* e^{i\Phi}\right)}{\text{Im}\left[\left(1 + m_1 m_2^*\right)e^{i\Phi}\right]}\nabla\times\mathbf{p}_\parallel^m. \tag{41}$$

Accordingly, for the whole ES of the field including the electric and magnetic parts of (38) – (40)

$$\mathbf{s}_E = \mathbf{s}_E^e + \mathbf{s}_E^m = \frac{1}{k^2}\frac{\text{Im}\left[\left(\frac{\tilde{\varepsilon}}{\varepsilon} + \frac{\tilde{\mu}}{\mu}m_1 m_2^*\right)e^{i\Phi}\right]}{\left(\frac{\tilde{\varepsilon}}{\varepsilon} + \frac{\tilde{\mu}}{\mu}\right)\text{Im}\left[\left(1 + m_1 m_2^*\right)e^{i\Phi}\right]}\nabla\times\left(\mathbf{p}_\parallel^e + \mathbf{p}_\parallel^m\right). \tag{42}$$

The result (41), (42) reduces to the simple rule of (31) in two cases: (i) for arbitrary waves if the dispersion is absent and (ii) for arbitrary dispersion if the waves' polarizations obey the condition $m_1 m_2^* = 1$. This condition is realized, e.g., if the beams possess the same circular ($m_1 = m_2 = i$) or 45º ($m_1 = m_2 = \pm 1$) polarizations. Note also that the longitudinal momentum in the right-hand side of the spin–momentum rule (42) includes the orbital momentum and the helicity-dependent part of the SM (see (39), (40)).

## 4. Surface plasmon-polariton

The next example involves surface plasmon-polariton (SPP) waves [7,11,28] which can be excited near the interface between the dielectric and conductive materials. In the simple model of Fig. 2, the two homogeneous media are separated by the plane interface $x = 0$; medium 1 ($x > 0$) is dielectric with electric and magnetic constants $\varepsilon_1$ and $\mu_1$, medium 2 ($x < 0$) is conductive and characterized by the frequency-dependent $\varepsilon_2(\omega)$ and $\mu_2(\omega)$; for simplicity, all permittivities and permeabilities are supposed to be real (this corresponds to negligible energy dissipation). In such a system, a double-evanescent TM wave may propagate along the interface with electric and magnetic field distributions [11,20,28,29]

$$\mathbf{E}_1 = \frac{A_s}{\varepsilon_1}\left(\mathbf{e}_x - i\frac{\kappa_1}{k_s}\mathbf{e}_z\right)\exp\left(ik_s z - \kappa_1 x\right), \quad \mathbf{H}_1 = \mathbf{e}_y A_s\frac{k_0}{k_s}\exp\left(ik_s z - \kappa_1 x\right), \quad (x > 0); \tag{43}$$

$$\mathbf{E}_2 = \frac{A_s}{\varepsilon_2}\left(\mathbf{e}_x + i\frac{\kappa_2}{k_s}\mathbf{e}_z\right)\exp\left(ik_s z + \kappa_2 x\right), \quad \mathbf{H}_2 = \mathbf{e}_y A_s\frac{k_0}{k_s}\exp\left(ik_s z + \kappa_2 x\right), \quad (x < 0). \tag{44}$$

where $k_0 = \omega/c$, $A_s$ is the coordinate-independent normalization constant, and

$$k_s^2 = \frac{\varepsilon_1\varepsilon_2\left(\varepsilon_1\mu_2 - \varepsilon_2\mu_1\right)}{\varepsilon_1^2 - \varepsilon_2^2}k_0^2, \quad \kappa_{1,2}^2 = \varepsilon_{1,2}^2 k^2\frac{\varepsilon_2\mu_2 - \varepsilon_1\mu_1}{\varepsilon_1^2 - \varepsilon_2^2}, \quad \frac{\kappa_1}{\kappa_2} = -\frac{\varepsilon_1}{\varepsilon_2}. \tag{45}$$

As usual, we accept that the medium 1 is a "normal" dielectric with $\varepsilon_1 > 0$, $\mu_1 > 0$; then, one of the two sets of conditions must hold to enable the TM SPP propagation: either

$$\varepsilon_2 < -\varepsilon_1, \quad \mu_2 > -\frac{\varepsilon_1\mu_1}{|\varepsilon_2|}$$

or

$$-\varepsilon_1 < \varepsilon_2 < 0, \quad \mu_2 < -\frac{\varepsilon_1\mu_1}{|\varepsilon_2|}.$$

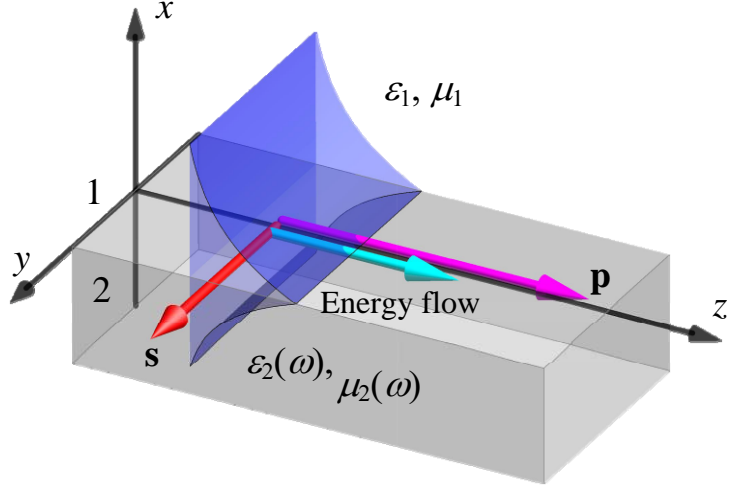

Fig. 2. Geometrical configuration of a system supporting the SPP propagation; the arrows show: (cyan) the energy flow (4); (magenta) momentum (7), (46); (red) spin (8), (47).

The dynamical characteristics of an SPP propagating in the system of Fig. 2 are calculated in Ref. [29]. The TM SPP represents a "classical" ES situation where the whole momentum is longitudinal and the whole spin is extraordinary. The volume part of the momentum is described by equations (see Eq. (32) of [29])

$$\mathbf{p} = \mathbf{p}_{\parallel} = \mathbf{e}_z g \frac{|A_s|^2}{2c} \frac{k_0}{k_s} \begin{cases} 2\mu_1 e^{-2\kappa_1 x}, & x > 0; \\ \dfrac{\tilde{\varepsilon}_2\mu_2 + \varepsilon_2\tilde{\mu}_2}{\varepsilon_2} e^{2\kappa_2 x}, & x < 0. \end{cases} \tag{46}$$

Calculating $\nabla \times \mathbf{p} = -\mathbf{e}_y \dfrac{\partial p_z}{\partial x}$ and comparing the results with the ES expression (Eq. (23) of [29]), we easily find that in this case, spin–momentum equation (2) accepts the form

$$\mathbf{s}_E = \mathbf{e}_y g \frac{|A_s|^2}{\omega k_s} \begin{cases} \dfrac{\kappa_1}{\varepsilon_1} e^{-2\kappa_1 x}, & x > 0 \\ -\dfrac{\tilde{\varepsilon}_2}{\varepsilon_2}\dfrac{\kappa_2}{\varepsilon_2} e^{2\kappa_2 x}, & x < 0 \end{cases} = \begin{cases} \dfrac{1}{2k_1^2}\nabla \times \mathbf{p}_{\parallel}, & x > 0; \\ \dfrac{1}{2k_2^2}\dfrac{2\mu_2\tilde{\varepsilon}_2}{\mu_2\tilde{\varepsilon}_2 + \tilde{\mu}_2\varepsilon_2}\nabla \times \mathbf{p}_{\parallel}, & x < 0. \end{cases} \tag{47}$$

where $k_i^2 = k_0^2\varepsilon_i\mu_i = k_0^2 n_i^2$ ($i = 1, 2$). Again, like in Section 3 (Eqs. (41), (42)), the result complies with the general rule (31) in case of negligible dispersion (however, this case is non-physical for the SPP fields). Note also that the specific form of the spin–momentum relation (47) in the lower medium ($x < 0$) agrees with the vanishing magnetic spin contribution of the TM SPP modes.

## 5. Discussion and conclusion

Simple examples, considered in the preceding sections, shed new light on the "spin–momentum equation" (2) introduced previously [24]. On the one hand, they distinctly confirm its universal character and wide applicability; on the other hand, its specific realizations (31), (42), (47) enable to more accurately determine the limits of its validity and special meanings of the "spin" and "momentum" terms united by (2). Additionally, Eqs. (42) and (47) demonstrate that the proportionality coefficient, uniting the spin density and the curl of momentum, depends on the medium dispersion and, as in (42), on the light field structure.

The first general conclusion is that the "generic" electromagnetic spin (8) does not obey Eq. (2): this can be easily seen if Eq. (2) is applied to the circularly polarized plane wave. However, in structured fields, it is reasonable to introduce the spin decomposition into the "ordinary" helicity-dependent and "extraordinary" helicity-independent parts. We do not develop a regular procedure for such a decomposition; moreover, in general case, it seems to be impossible or physically meaningless. At the same time, in some important situations discussed above, it makes sense, and it is this polarization-independent spin that enters the spin–momentum rule (2).

Second conclusion concerns the specification of the momentum term in Eq. (2). It, again, cannot be the "generic" field momentum (4), nor its spin (6) or orbital (7) part, but it is the properly defined longitudinal momentum (and then the ES of Eq. (2) appears as the transverse spin [11,12]). In all the considered examples, the well-defined longitudinal direction exists, so the longitudinal momentum can be easily identified for paraxial beams (Section 2) or in the interference field (Section 3). As to the SPP field of Section 4, there the whole field momentum is longitudinal, and that is why this case appears especially "natural" for the "spin–momentum locking" effects [21–24]. It should be noted, however, that the requirement for $\mathbf{p}$ in Eq. (2) to be associated with the longitudinal direction is not absolute; possibly, other situations without such association can exist.

From the mathematical point of view, Eq. (2) as well as its specific versions (31), (42) and (47) establishes the formal association of the helicity-independent ES and the curl of the longitudinal momentum density. Due to this form, the "extraction" of the ES from the "full" optical spin resembles the separation of the SM (8) from the general electromagnetic momentum: the spin density serves a vector-potential for the SM, and, quite similarly, the longitudinal momentum density is a vector-potential for the ES. It might be expected that the formal analogy of Eqs. (1) and (2) testifies for the certain reciprocity in their nature: the ES appears due to the inhomogeneity of the momentum (energy flow) like the SM emerges due to inhomogeneity of the spin (helicity).

Also, the mathematical form of Eqs. (2) and (31), (42), (47) invokes clear analogies with the motions in a continuous medium. Indeed, rotational properties of a hydrodynamic flow with the inhomogeneous velocity distribution $\mathbf{v(R)}$ is characterized by the vorticity [25]

$$\mathbf{\Omega} = \frac{1}{2}\nabla \times \mathbf{v} \qquad (48)$$

determined quite similarly to the ES (2) and its special cases (31), (42) and (47). Notably, the non-zero vorticity may exist in flows without visible rotational motion of the fluid particles. For example, in a layered flow near the plane wall, a one-dimensional velocity distribution $\mathbf{v} = \mathbf{e}_z v_z(x)$ appears with the vorticity $\mathbf{\Omega} = -(1/2)\mathbf{e}_y(\partial v_z / \partial x)$, and this transparently hints at the nature of the helicity-independent spin.

The analogy between Eqs. (48) and (2) is not only formal: the layered material flow in fluid and the layered distribution of the electromagnetic momentum in the light field both mean the layered distribution of the energy transport. In this context, the resemblance between the momentum $\mathbf{p}_\parallel$ (see Eqs. (31), (42), (47)) and the velocity $\mathbf{v}$ becomes rather meaningful, and the helicity-independent spin emerges as a sort of the "energy flow vorticity" orthogonal to the main propagation direction. The picture of the "optical" layered flow is especially spectacular in case of the SPP where, due to the one-dimensional $\mathbf{p}_\parallel(x)$ distribution, Eq. (48) immediately yields $\mathbf{s}_E \propto -\mathbf{e}_y(\partial p_\parallel / \partial x)$. The non-zero vorticity of the layered flows testifies for the presence of a "hidden" rotations in them; in this aspect, the "optical" situation looks more physically transparent because, in contrast to the mechanical flows, the origin of the "hidden" rotations is obvious: instantaneous electric and magnetic vectors rotate in the longitudinal plane [7,11–15]. Hence one may conclude that the spin–momentum equations similar to Eq. (2) are typical for the "photonic wheels" [14,15] and can be considered as their natural formal attributes.

Finally, the main outcome of this study can be summarized as follows: The spin–momentum law (2), with some variations of the proportionality coefficient, takes place whenever the well defined longitudinal and transverse directions exist in the field, and the longitudinal momentum is inhomogeneous over the cross section. Then the helicity-independent spin appears within the cross section, and Eq. (2) describes its relation with the longitudinal component of the field momentum. This picture of the ES emergence can probably be extrapolated to other wave fields, first of all, acoustic [17–19], where the spin–momentum relations similar to Eq. (2) will be also relevant.

**Acknowledgements**

The authors are grateful to K. Bliokh (RIKEN, Japan) for fruitful discussion.